\documentclass{ws-ijmpd}

\begin{document}

\markboth{I.B.Khriplovich and D.L.Shepelyansky}
{Capture of dark matter by the Solar System}

%
\catchline{}{}{}{}{}
%

\title{CAPTURE OF DARK MATTER BY THE SOLAR SYSTEM }

\author{I.B.KHRIPLOVICH}

\address{Budker Institute of Nuclear Physics, 630090 Novosibirsk, Russia\\
I.B.Khriplovich(at)inp.nsk.su}

\author{D.L.SHEPELYANSKY}

\address{Universit\'e de Toulouse, UPS,
Laboratoire de Physique Th\'eorique (IRSAMC)\\
F-31062 Toulouse, France;\\
CNRS, LPT (IRSAMC), F-31062 Toulouse, France\\
dima(at)irsamc.ups-tlse.fr}

\maketitle

\begin{history}
\received{30 June 2009}
\revised{XX XXXX 2009}
\comby{Managing Editor}
\end{history}

\begin{abstract}
We study the capture of galactic dark matter by the Solar System.
The effect is due to the gravitational three-body interaction
between the Sun, one of the planets, and a dark matter particle.
The analytical estimate for the capture cross-section is derived and
the upper and lower bounds for 
the total mass of the captured dark matter particles are found. The
estimates for their density are less reliable.
The most optimistic of them give an enhancement 
of dark matter density  by about three orders of magnitudes  compared to its
value in our Galaxy. However,  even this optimistic
value remains below the best
present observational upper limits by about two orders of
magnitude.
\end{abstract}


\section{Introduction}	

The Dark Matter (DM) density in our Galaxy is about 
(see e.g. Ref.~\refcite{ber}):
\begin{equation}
\label{rho}
\rho_g \simeq 4 \cdot 10^{-25}\;\rm{g/cm}^3\,.
\end{equation}
However, only upper limits on the level of $10^{-19}\;
\rm{g/cm^3}$ (see below) are known for the density of Dark Matter
Particles (DMP) in the Solar System (SS). Meanwhile, information
on their density is of great importance for the experiments aimed
at the detection of DM.

The question of capture of Weakly Interacting Massive Particles (WIMP)
or DMP by the SS was pioneered in Refs.~\refcite{ga,le}. 
Very recently these studies are
pushed further by extensive numerical simulations performed in 
Refs.~\refcite{ap,ap1}. Our interest to this problem was
attracted by a recent Ref.~\refcite{xs} where an estimate 
is given for the DM density
in SS, as resulting from the gravitational capture of galactic
DMP's. According to the conclusions of Ref.~\refcite{xs}, the density of
the captured DM, for instance, at the Earth orbit is about
$10^{-20}\; \rm{g/cm^3}$, only about an order of magnitude below
the best upper limits on it.

In the present note we perform new analytical analysis of the gravitational
capture of galactic DMP's by the SS. According to our results, the
increase of the DM density in the SS due to this capture is small,
certainly well below $10^{-20}\; \rm{g/cm^3}$.

\section{Dimensional estimate for mass of captured dark matter}

The Solar System is immersed in the halo of dark matter
and moves together with it around the center of our Galaxy. To
simplify the estimates, we assume that the Sun is at rest with
respect to the halo. The dark matter particles in the halo
are assumed to have in the reference frame, co-moving with the
halo, the Maxwell distribution (see Ref.~\refcite{ag}):
\begin{equation}
\label{ma}
f(v)\,dv = \sqrt{\frac{54}{\pi}}\,\frac{v^2 dv}{u^3}
\exp{\left(-\frac{3}{2}\frac{v^2}{u^2}\right)}\,,
\end{equation}
with the local rms velocity $u \simeq 220$ km/s.

Let us elucidate what looks to be the most efficient mechanism of
the DMP capture. It was pointed out and partly analyzed (though
for the capture of comets, but not of DMP's), by Petrosky in Ref.~\refcite{pe}
and Chirikov and Vecheslavov in Ref.~\refcite{cv}. Of
course, a particle cannot be captured by the Sun alone. The
interaction with a planet is necessary for it, this is essentially
a three-body problem of the Sun, planet and DMP. Obviously, 
the capture is dominated by the
particles with orbits close to
parabolic ones with respect to the Sun, and with the distances 
between their perihelia and the Sun comparable with
the radius of planet orbit $r_p$.

The capture can be effectively described by the so-called
restricted three-body problem (see for instance Refs.~\refcite{sz,3b}). In
this approach the interaction between two heavy bodies (the Sun
and a planet in our case) is treated exactly. As exactly is
treated the motion of the third, light body (a DMP in our case) in
the gravitational field of the two heavy ones. One neglects
however the back reaction of a light particle upon the motion of
the two heavy bodies. Obviously, this approximation is fully
legitimate for our purpose. Still, the restricted three-body
problem is rather complicated, and requires in the present case
both subtle analytical treatment and serious numerical
calculations (see Ref.~\refcite{pe}). Under certain conditions the dynamics of light
particle (e.g. DPM) becomes chaotic.

However, the amount of DM captured by the SS can be found by means
of simple estimates. The total mass captured by the Sun (its mass
is $M$) together with a planet with mass $m_p$, during the
lifetime
\begin{equation}
\label{T}
T \simeq
4.5 \cdot 10^9 \;\rm{years} \simeq   10^{17}\; \rm{s}
\end{equation}
of the SS, can be written as follows:
\begin{equation}
\label{de}
\Delta m_p = \rho_g T <\sigma v>\,;
\end{equation}
here $\sigma$ is the capture
cross-section. The product $\sigma v$ is averaged over
distribution (\ref{ma}); with all typical velocities in the SS
much smaller than $u$, this distribution simplifies to
\begin{equation}
\label{rj}
f(v)\,dv = \sqrt{\frac{54}{\pi}}\,\frac{v^2 dv}{u^3}\,.
\end{equation}

To estimate the average value $<\sigma v>$, we resort to
dimensional arguments, supplemented by two rather obvious physical
requirements: the masses $m_p$ and $M$ of the two heavy components
of our restricted three-body problem should enter the result
symmetrically, and the mass $m_d$ of the light component (DMP) should not
enter the result at all. Thus, we arrive at
\begin{equation}
\label{av}
<\sigma v> \,\sim \sqrt{54\pi}\;\;\frac{k^2\,m_p\,M}{u^3}\,;
\end{equation}
here $k$ is the Newton gravitation constant; an extra power
of $\pi$, inserted into this expression, is perhaps inherent in $\sigma$.
The final estimate for the captured mass is
\begin{equation}
\label{de1}
\Delta m_p \sim \rho_g T \sqrt{54 \pi}\;\;\frac{k^2\,m_p\,M}{u^3}\,.
\end{equation}
Since the capture would be impossible if the planet were not bound
to the Sun, it is only natural that the result is proportional to
the corresponding effective "coupling constant" $k m_p M$.

Thus obtained values for the masses of DM captured due to the
planets of the SS, are presented in Table 1. We quote also therein
the corresponding results of Ref.~\refcite{xs} for these masses. The
disagreement is huge for all planets, and especially for the light
ones where it exceeds two orders of magnitude. We cannot spot
exactly its origin since the calculations of Ref.~\refcite{xs} involve
rather complex numerical simulations (it is possible that their 
assumption of  capture radius $r_b \sim r_p(m_p/M)^{1/3}$ 
does not correspond to reality).
On the other hand, however, we cannot see any
reasonable possibility for a serious increase of our results.
Moreover, in a sense they can be considered as upper limits for
the amount of the captured DM, at least because we have neglected
here the inverse process, that of the ejection of a captured DMP
due to the same three-body gravitational interaction.
The result (\ref{de1}) is given for the three body problem.
The dynamical mechanism of capture is described below
in next Section.
The contribution of the diffusive (non-dark) matter
in the SS should be significantly smaller since 
a homogeneously distributed dust gives compensation of
gravitational forces acting on DMP.

The total mass $\Delta m_T$ of the DM captured 
by the planets is strongly
dominated by the heavy Jovian planets, Jupiter, Saturn, Uranus,
Neptune, and constitutes according to Table 1 about 
$\Delta m_T \sim 1.5 \cdot 10^{21}$~g. 
This value is small as compared to the total mass 
$\sim 10^{33}$ g of the common matter in the SS. It is small even as
compared to the total non-captured mass of the DM in the SS: 
this total mass,
calculated with value (\ref{rho}) for the DM density, constitutes
$\sim 10^{31}$~g (we assume here that the effective radius of the
SS is about $ 10^5$~au). However, it is 
an order of magnitude larger than the DM mass of density (\ref{rho}) 
inside the radius of Neptune orbit $r_N \approx 30$ au.

The contribution to the discussed effect of the
diffuse (non-dark) matter in the SS should 
be significantly smaller since in a homogeneous dust
the gravitational forces acting on DMP are compensated.

The dynamical mechanism of capture is described in the next Section.

\begin{table}[ph]
\tbl{DM mass captured by planets (in $g$)}
{\begin{tabular}{@{}ccccccccc@{}} \toprule
   Planet & Mercury & Venus & Earth & Mars
     & Jupiter &  Saturn & Uranus & Neptune \\
 \colrule
 this work & 0.22$\cdot10^{18}$ & 3.2$\cdot10^{18}$ & 3.9$\cdot10^{18}$ &  0.42$\cdot10^{18}$ &  1239$\cdot10^{18}$ & 372$\cdot10^{18}$  &  57$\cdot10^{18}$ &  67$\cdot10^{18}$  \\
 Xu, Siegel Ref.~\refcite{xs}  & 0.42$\cdot10^{20}$ & 3.5$\cdot10^{20}$ &
 3.8$\cdot10^{20}$ &  1.2$\cdot10^{20}$  & 49$\cdot10^{20}$ &
 28$\cdot10^{20}$ &  12$\cdot10^{20}$ & 16$\cdot10^{20}$   \\
\botrule
\end{tabular} \label{ta1}}
\end{table}

\section{Dynamical approach}

For the restricted three-body problem,
in a close similarity to dynamics of comets (see Refs.~\refcite{pe,cv}), 
the DMP dynamics can be described
by a symplectic area-preserving map
\begin{equation}
\label{map}
\bar{w} = w + F(\phi) \; \; , \;\;
\bar{\phi}=\phi+ 2\pi \bar{w}^{-3/2} \; .
\end{equation}
Here $w=-2E r_p/km_dM$ is the energy of DMP with mass $m_d$
rescaled by its gravitational energy on distance $r_p$ from the Sun;
$\phi$ is the phase of the planet on its circular orbit
at the moment when the DMP is at the perihelion and $F(\phi)$ 
is a certain periodic function of $\phi$.
Bars denote the new values of variables after one rotation
around the Sun. The physical meaning of this dynamical map
is rather simple: the first equation gives the change of DMP energy
after one passage near the Sun, the second equation
gives the change of the planetary phase between two passages of DMP
and is essentially determined by the Kepler law. The first equation is
valid also for scattering particles with positive energy ($w<0$).
Thus DMP can be captured by the Sun and the planet only
if its rescaled energy $|w|<F_{max}$. After the capture, the DMP 
dynamics is described by map (\ref{map}) until ejection.
To compute the captured DM mass $\Delta m_p$ we assume that,
after being once captured, the DMP remains captured for the whole
life time $T$ of SS. In this way we obtain the maximum bound
for $\Delta m_p$.

The kick function $F(\phi)$ was computed in Ref.~\refcite{pe}
for the case when a comet (or DMP) and a planet 
move in one plane and when the perihelion distance $q>r_p$.
In this case $F(\phi) =  (m_p/M) \beta(q/r_p) \sin \phi$
and the function
$\beta(x) \approx 26 \exp(-4x^{3/2}/3\sqrt{2})/x^{1/4}$
so that $\beta(1) \approx 10$. Effectively the function $F$
is determined by the frequency Fourier component of the force
between the planet and DMP, since the rotation of the planet is
rapid compared to the rotation of DMP the amplitude of the component
is exponentially small for $q \gg r_p$ 
when the DMP motion is smooth and
analytical. In this case $\beta$ is exponentially small
and there is practically no trapping of DMP. For $q \sim r_p$
the motion is not analytic due to close passage between
the planet and DMP and $\beta$ is relatively large. 
In this case a DMP with rescaled 
energies $-w < \beta m_p/M$ can be captured by the planet.
It is interesting to note that 
the map (\ref{map}) with  $F(\phi) \sim \sin \phi$
is known as the Kepler map. It describes
the process of microwave ionization of Rydberg
atoms and chaotic autoionization of molecular Rydberg states
(see Ref.~\refcite{be} and Refs. therein).

We note that the energy change of DMP given by $F(\phi)$
results from the integration over the whole orbit
rotation of DMP around the Sun which includes many orbital
periods of the planet. Thus this energy change appears
from long-range interaction and has qualitatively different
origin compared to local close collisions between
DMP and planet which were assumed to give the main contribution
for DMP energy change in Refs.~\refcite{ga,le,xs}.

Let us now estimate the capture cross-section $\sigma$
assuming that for all DMP the dynamics is described by the Kepler map
with fixed $\beta \sim 1$, Then only DMP with energies
$|w| = v^2 r_p/k m_p M =v^2/v_p^2 < \beta m_p/M$ are captured 
under the condition that 
$q < r_p$ (here $v_p$ is the velocity of the planet).
The value of $q$ can be expressed via the 
DMP parameters at infinity where its velocity is $v$ and 
its impact parameter is $r_d$ and hence
$q=(v r_d)^2/2 k M$ (see Ref.~\refcite{la}).
Since $q \sim r_p$ we obtain the cross-section 
\begin{equation}
\label{sigma}
\sigma \sim \pi r_d^2 \sim  2\pi k M r_p/v^2 \sim
2 \pi r_p^2 (v_p/v)^2 \sim 2\pi  r_p^2 M/(\beta m_p) \; ,
\end{equation}
where the last relation is taken for those typical 
velocities $v^2 \sim \beta v_p^2 m_p/M$
at which the capture of DMP takes place
(for $q \approx 1.4 r_p$ we have $\beta \approx5$). 
Then Eqs. (\ref{de}), (\ref{sigma})
give the captured mass $\Delta m_p$ of (\ref{de1}) with 
an additional numerical factor  $\beta \sim 1$.

According the above estimates DPM captured by Jupiter
have typical velocities at infinity 
$v \sim (\beta m_p/M)^{1/2} v_p \sim 1 km/s$
for typical $\beta \sim 5$ and $m_p/M \approx 10^{-3}$,
$v_p \approx 13 km/s$. This value of $v$ is in
a good agreement with the numerical simulations of Peter Ref.~\refcite{ap1}
which give typical captured DPM velocities 
for Jupiter of $1 km/s$.

\begin{table}[ph]
\tbl{Density of DM for planets (in $g/cm^3$)}
{\begin{tabular}{@{}ccccccc@{}} \toprule
   Planet & Mercury & Venus & Earth & Mars
     & Jupiter &  Saturn  \\
 \colrule
$\Delta \rho_p$ &2.7$\cdot10^{-22}$ & 6.0$\cdot10^{-22}$ &
2.7$\cdot10^{-22}$ & 8.4$\cdot10^{-24}$ & 6.2$\cdot10^{-22}$ &
3.0$\cdot10^{-23}$  \\
 \colrule
$\rho_{\rm DM}$ &1.8$\cdot10^{-21}$ & 1.5$\cdot10^{-21}$ &
9.3$\cdot10^{-22}$ & 6.6$\cdot10^{-22}$ & 6.5$\cdot10^{-22}$ &
3.0$\cdot10^{-23}$  \\
\botrule
\end{tabular} \label{ta2}}
\end{table}

\begin{table}[ph]
\tbl{Angle of perihelia precession (in  seconds per century)}
{\begin{tabular}{@{}cccc@{}} \toprule
   Planet &  Venus & Earth & Mars  \\
 \colrule
$\delta \phi_{\rm th}$ & 8.6248 & 3.8388 & 1.3510 \\
 \colrule
$\delta \phi_{\rm obs}$  & 8.6247 $\pm$ 0.0002 & 3.8390 $\pm$ 0.0003 & 1.3512 $\pm$ 0.0003 \\
\botrule
\end{tabular} \label{ta3}}
\end{table}

Another interesting feature of the analytical expression for the
cross-section of captured particles $\sigma$ (\ref{sigma})
is that it is much larger than the area of the planet orbit.
In fact, $\sigma$ diverges at small velocities as
$\sigma \sim 1/v^2$ but this divergence is weaker
compared to the divergence of the Reserford cross-section.
In our case of the restricted three-body problem
the divergence appears due to the property
of the Kepler motion where the DMP distance at perihelion
is proportional to square of the orbital momentum
which in its turn is proportional to the product 
of the velocity $v$ and impact parameter $r_d$ 
at infinity. In addition it is important
to use the  value of typical DMP velocity captured 
by the planet for perihelion distance of the order
of $r_p$. This leads to the analytical equation (\ref{sigma})
for the capture cross-section in the restricted 
three-body problem.

\section{Density of dark matter}

 While the total masses $\Delta m_p$ of the captured DM can be
(hopefully) described by simple dimensional estimate (\ref{de1}),
the situation for the corresponding DM densities $\Delta \rho_p$ is
more subtle. The reason is as follows. The captured DMP's had
initial trajectories predominantly close to parabolas with respect
to the Sun, and the velocities of these DMP's change only slightly
as a result of scattering. Therefore, it is quite natural that the
final, elliptical trajectories of these DMP's have large
semi-major axes. 

Indeed, DMP captured into  an elliptic trajectory had initially a
hyperbolic trajectory, focussed at the Sun and close to a
parabolic one. As a result of the capture, the eccentricity $e$ of
the trajectory changes from $e = 1+ \epsilon_1$ to $e = 1-\epsilon_2$, with
$\epsilon =\epsilon_1 + \epsilon_2 \ll 1$. 
It is quite natural that the final,
elliptical trajectories of the captured DMP's have large
semi-major axes.

To estimate their typical values, we recall 
(see Ref.~\refcite{la}) that the radius-vector
$r$ of a captured DMP (counted off the Sun) is related to the
azimuthal angle $\phi$ as follows:
\begin{equation}
r = p\,/(1 + e \cos \phi)\,,
\end{equation}
where $p$ is the so-called orbit parameter (its value is
irrelevant for our line of reasoning). Obviously, the maximal $r_{max}$ and 
minimal distance $r_{min}$ from the Sun
correspond to $\cos \phi = \pm 1$, so that their ratio is
\begin{equation}
r_{max}/r_{min} = (1 + e)/(1 - e)\,.
\end{equation}
In the numerator of this ratio, we can safely put with our 
accuracy $1 + e \simeq 2$. As to its denominator, we recall that
the difference $1 - e$ is related to the gravitational
perturbation by planet, and therefore is proportional to $m_p$.
Thus, by dimensional reasons,
\begin{equation}
r_{max}/r_{min} \sim M/m_p\,.
\end{equation}

The minimal distance between DMP and the Sun $r_{min}$
should be on the same order of magnitude as the
radius $r_p$ of the planet orbit. Therefore, the semi-major axis
$a$ of resulting ellipse is huge:
\begin{equation}
 r_{max} \sim r_p \,(M/m_p) \; .
\end{equation}
In particular, in the case of Jupiter our estimate gives
$r_a \sim 10^3 r_p$. A similar numerical factor appears
in Ref.~\refcite{pe}.

According to the numerical calculations 
of Ref.~\refcite{pe}, in the case of Jupiter the values of the semi-major
axes $r_a$ for the resulting trajectories belong to the interval
10$^3$ --- 10$^4$ au for $q/r_p=4$ --- $6$. 
The fact that it is comets that are considered in
Ref.~\refcite{pe}, but not DMP's, is obviously of no importance for
this conclusion. The minimum value of $r_a$ is
defined by the maximum $w_{ch} \sim r_p/r_a$ value 
which can be reach by an injected DMP
during its chaotic motion.
In fact $w_{ch}$ is the chaos border and according to
the Chirikov resonance-overlap criterion (see Ref.~\refcite{ch})
we have $w_{ch} \approx (3 \pi \beta (m_p/M))^{2/5}$
as it was shown in Ref.~\refcite{pe}.
For Jupiter $m_p/M \approx 10^{-3}$ and at $\beta \approx 5$
corresponding to $q \approx 1.4 r_p$
we have $r_a/r_p \approx 1/w_{ch} \approx 3$.
We note that this value of $\beta$ gives the maximum $F_{max} \approx 0.005$
corresponding to the similar value found 
for the comet Halley (see Ref.~\refcite{cv}). 
In fact, the data presented in Ref.~\refcite{cv}
show that the comet Halley have the chaos border 
around $w_{ch} \approx r_p/r_a \approx 0.3 $ (see Fig.3 
in Ref.~\refcite{cv}).

Of course the values of $r_a$ linked to the chaos border in $w$
are the minimum ones since during its chaotic dynamics DMP
have also $0< w \ll w_{ch}$ with larger $r_a$. However,
we are interested in orbits captured for very large times $T$ (\ref{T}).
Such times are by two orders of magnitude larger 
than a typical diffusive life time
of comet Halley found to be of the order of $10^7$ years (see Ref.~\refcite{cv}). 
It is known that chaotic trajectories may be sticking to boundaries of 
integrable islands for very long times (see Ref.~\refcite{cs} and Refs. therein)
and hence we can expect that those orbits will be somewhere in vicinity
of the chaos border around $w_{ch} \sim 1$ with $r_a \sim r_p$.
In fact, for the case when the inclination angle 
between the planes of DMP  and planet  $\theta_i > 0$ and when
$q < r_p$ the function $F(\phi)$ contains higher harmonics
of $\phi$ (see the case of the comet Halley in Ref.~\refcite{cv}). 
This leads to easier emergence of chaos
so that even for light planets one may have
the chaos border $w_{ch} \sim 1$.

Therefore we  can make an assumption resulting in the most optimistic
prediction for the  "partial" dark matter densities $\Delta \rho_p\,$.
We assume that each of the total masses
$\Delta m_p$ of the captured DM occupies the volume $(4\pi/3) r_p^3$
where $r_p$ is the orbit radius of the corresponding planet. We do
not claim that this assumption is correct, but believe, however,
that the comparison of its (almost certainly, overoptimistic)
results with the observational limits will be instructive. The
corresponding values of the "partial" dark matter densities 
$\Delta \rho_p =\Delta m_p/(4\pi r_p^2/3)$ (in
g/cm$^3$) are presented in Table 2. We omit in it the densities
due to Uranus and Neptune, tiny even at the discussed scale. Then,
in accordance with the accepted model, the total dark matter
density $\rho_{\rm DM}$ at a given radius does not coincide with the
corresponding $\Delta \rho_p$. It includes, in line with it, the
sum of the contributions to the density due to all the planets,
outer with respect to the given one.

\section{Ergodic time scale}

The estimates given above neglect ejection of DMP
from the SS. Such an assumption is not justified
if the DMP dynamics in SS becomes completely ergodic
on a time scale $T_e \ll T$. Then after the time $T_e$
the detailed balance principle becomes valid
and the density of captured DMP becomes 
as its galactic density as it was argued in Ref.~\refcite{ga}
(see also discussion in Refs.~\refcite{le,ap,ap1}). However, the estimate of $T_e$
have certain subtle points. In the frame
of the map (\ref{map}) it is given by 
the  diffusion time from $w =0$ to $w=w_{ch}$. For the Kepler map
the diffusion coefficient is $D \approx \beta^2 (m_p/M)^2/2$
and hence $T_e \sim 2(M w_{ch}/\beta m_p)^2 T_d$ where
$T_d$ is an average period of DMP.
For the case of  Jupiter such an estimate gives a satisfactory
value of $T_e \sim 10^7$ years 
for the case of the comet Halley as it is discussed in 
Ref.~\refcite{cv}. However, $\beta$ sharply decreases with the increase
of perihelion distance $q \propto \ell^2$,
where $\ell$ is the orbital momentum of DMP.
As a result a growth of $\ell$ can give
sharp increase of $T_e$ which can become comparable with $T$. 
The  effects linked to variations of $\ell$ were not 
considered in Ref.~\refcite{cv}. They are properly treated in the numerical
simulations of Refs.~\refcite{ap,ap1}
but there only the effect of Jupiter is considered.
Other planets and fluctuating galactic fields
can give stronger growth of 
$\ell$ with a significant increase of $T_e$.
Indeed, from the studies of Rydberg atoms in a microwave
field it is known that time-oscillating 
space homogeneous fields can produce
strong variation of eccentricity of orbits
(see Ref.~\refcite{cgs} and Fig.~11 there).
Also it is known that noise generates penetration
of chaotic trajectories inside integrable islands
and very slow decay of Poincar\'e recurrences
with diverging trapping time (see Ref.~\refcite{cs84}).
In addition to that recent large scale numerical simulations 
of Ref.~\refcite{las}
show significant changes of eccentricity of planets
on a time scale of the order of $T$.
Therefore the question about $T_e$ value for DMP captured 
by the SS requires further studies. It is not excluded that 
it is comparable or even larger than $T$. In such a situation 
the upper bound (\ref{de1}) will be close to the real value
of total captured mass.

In any case it is clear that there is practically no ejections
of captured particles on a time scale of DMP orbital period $T_c$.
A typical captured DMP rescaled energy is $w_c \sim \beta m_p/M$
corresponding to one iteration of the map (\ref{map})
which gives a change of $w$ from negative to positive values.
The rotation period of such DMP is rather large
compared to the period $T_p$ of the planet: 
$T_c/T_p \sim (\beta m_p/M)^{-3/2}$. For the case of Jupiter
$T_p \approx 11$ years and at $\beta \sim 1$ we have 
$T_c \sim 3 \cdot 10^4 T_p \sim  3 \cdot 10^5$ years.
This is by  a factor $10^4$ shorter than the SS life time $T$.
This gives the lower bound of the captured DM mass
which is obtained by replacing $T$ by $T_c$ in the equation
(\ref{de1}).

\section{Observational upper limit on the density of dark matter}

At last, let us consider the observational data on the DM in
our SS. The most reliable and accurate information on it follows
from the studies of the perihelion precession of Venus, Earth, and
Mars. Under the assumptions that the DM density $\rho_{\rm DM}$ 
is distributed spherically symmetric with respect to the Sun
and that the eccentricity of the planetary orbit is small, the
corresponding relative shift of the perihelion per period is
(see Ref.~\refcite{kh} and Refs. therein):
\begin{equation}
\frac{\delta \phi}{2\pi} = -\frac{2\pi \rho_{\rm DM} r^3}{M}\,,
\end{equation}
where $r$ is the radius of the orbit. This relation gets almost
obvious (up to an overall numerical factor) if one recalls that,
in virtue of the Gauss theorem, for a spherically symmetric
density $\rho(r)$ the action of the DM inside the orbit reduces to
that of a point-like mass, and therefore does not induce the
perihelion precession. On the other hand, for such density
$\rho(r)$, the DM outside the orbit does not influence at all the
motion of a planet.

The recent, most precise observational data (see Ref.~\refcite{pi}) on the precession of
perihelia are presented in Table 3 
(therein the theoretical values $\delta \phi_{\rm th}$
of the perihelion rotation and the results 
of observations $\delta \phi_{\rm obs}$ are
given in angular seconds per century). With these data, 
one arrives at the upper
limits on the DM density at the distances from the Sun, 
corresponding to the orbit
radii of Venus, Earth, and Mars, on the level of
\begin{equation}
\label{rdm}
\rho_{\rm DM} < 2 \cdot 10^{-19}\; \rm{g/cm^3}\,.
\end{equation}

This observational upper limit exceeds by about two
orders of magnitude the results (almost certainly overestimated)
presented in Table 2.

\section{Summary}
Our results do not mean, however, that the searches for the dark
matter in the SS are senseless. Of course, the capture of the
Galactic DM analyzed here is not the only conceivable source of
the dark matter in the Solar System. It is quite possible in
particular that the Solar System itself has arisen due to a local
high-density fluctuation of the dark matter.

Now on the related theoretical problems.
To obtain more firm results for the captured DM mass and density
one needs to take into account that the kick function 
$F(\phi)$ in the map (\ref{map})
depends on an inclination angle between planes of DMP and planet orbits
and also on the DMP perihelion distance. However, a typical case
of the comet Halley analyzed by Chirikov and Vecheslavov in Ref.~\refcite{cv} 
gives the map function of a 
form similar to  one discussed here so that the estimates presented 
should be also applicable for such more general DMP orbits.
Further analytical and numerical studies are required
for a better understanding of DMP
dynamics inside the Solar System.

\section*{Acknowledgments}

We thank J.~Edsj\"o and A.H.G.~Peter for useful remarks
and for pointing to us relevant works on WIMP and DMP capture by the SS.
The work of I.B. Khriplovich was supported in part by the Russian Foundation
for Basic Research through Grant No. 08-02-00960-a.


\end{document}